\documentclass[10pt,a4paper,twoside]{article}
\usepackage{amssymb}
\usepackage{vmargin,fancyheadings,multicol,ifthen,cite,
        epsfig,wrapfig,calc,dcolumn,doublespace}
\usepackage{wgn2}

\def\E{\mbox{E}}
\def\ZHR{\mbox{ZHR}}

\begin{document}

\begin{WGNpaper}{twocol}{
\sectiontitle{Meteor science}
\title{Estimating meteor rates using Bayesian inference}
\author{Geert Barentsen\footnote{Armagh Observatory, 
College Hill, Armagh BT61 9DG, U.K., {\tt gba@arm.ac.uk}}, 
Rainer Arlt\,\footnote{Leibniz-Institut f\"ur Astrophysik Potsdam, An der Sternwarte 16, 14482 Potsdam, Germany},
Hans-Erich Fr\"ohlich\,$^2$}


\abstract{
A method for estimating the true meteor rate $\lambda$ from a small number of
observed meteors $n$ is derived.
We employ Bayesian inference with a Poissonian likelihood function.
We discuss the choice of a suitable prior
and propose the adoption of Jeffreys prior, $P(\lambda)=\lambda^{-0.5}$, which yields
an expectation value $E(\lambda) = n+0.5$
for any $n\geq 0$.
We update the ZHR meteor activity formula accordingly, and explain how 68\%- and 95\%-confidence intervals can be computed.
}
}

\section{Introduction}
The formula commonly used to estimate the Zenithal Hourly Rate (ZHR) of meteors  
as given in the Handbook of the International Meteor Organization (IMO; Rendtel \& Arlt, 2008) and used in many meteor activity graphs (e.g., Arlt \& Barentsen, 2006), is given by:
\begin{equation}
\E(\ZHR) = \frac{ n_{\rm tot} + 1 } { T } \quad\mbox{with}\quad \sigma=\frac{ \sqrt{n_{\rm tot} + 1} } { T },
\label{zhr}
\end{equation}
where $n_{\rm tot}$ is the total number of meteors counted in a number of 
observing intervals ($n_{\rm tot} = \sum_{i} n_i$) and $T$ is the observing 
time weighted by a given correction factor for each interval  
($T = \sum_{i} T_{{\rm eff},i} / C_i $).

The use of $n_{\rm tot}+1$ rather than $n_{\rm tot}$ often surprises observers,
because it yields a ZHR which is larger than zero even when no meteors are observed.
Whilst Arlt (1999) already indicated that $n_{\rm tot}+1$ 
is used to account for the effects of small-number statistics,
this paper will explain the formula in more detail.
However, contrary to the earlier publications, we will suggest that
 $n_{\rm tot}+0.5$ rather than $n_{\rm tot}+1$ is the most appropriate formula.

We will first describe the problem of small-number statistics in \S2. 
We then describe the solution using Bayesian inference in \S3,
followed by a discussion on the choice of the prior assumptions in \S4. 
In \S5 we explain how confidence intervals may be computed,
and in \S6 we provide examples. 
Finally in \S7 we discuss the effect of 
correction factors and in \S8 we present the conclusions. 

\section{Problem description}
As explained by Bias (2011) in a recent issue of this journal,
the {\em observed\/} meteor rate often tends to
underestimate the {\em true\/} meteor rate due to small-number statistics. 
This may be understood using the following example: 
if we would attempt to estimate the frequency of winning the lottery based on a small group of 10 players, 
we are most likely to find that none of these players have ever won and that the winning frequency equals zero.
Of course the true probability of winning the lottery is slightly larger than zero, 
but the quantity cannot reliably be obtained by extrapolating from a small number of players.

An identical effect occurs when the ZHR is extrapolated from a low number of meteors.
Indeed, even when zero meteors are observed in an interval of finite length, 
the true average rate may well have been larger than zero because we know that the
interplanetary space is not empty. The observer might have been unlucky, or the 
interval might have been short with respect to the true rate. 
Such situation may occur even during major showers,
e.g. when computing rates for 1-minute intervals.
The ZHR formula must take this into account in order to 
produce reliable estimates in all situations.

\section{Solution: Bayesian inference}
A common technique used in statistics to estimate parameters in 
the presence of sparse data is called {\em Bayesian inference}.
In brief, one constructs a parameterized model which one thinks describes the source of the data. The probability of this
model to have produced the data is then computed for each possible set of
parameters, taking into account any known prior constraints on the parameters.
The resulting set of probabilities is called the posterior distribution,
from which expectation values and confidence intervals for the free parameters
may be derived.

In our case, the data is the observed meteor rate $n$ (in arbitrary time $T$).
Our only free model parameter is the true meteor rate $\lambda$ (i.e. ZHR).
There are many values of $\lambda$ which may explain a given $n$,
each having the conditional probability $P(\lambda|n)$. This notation means
the probability of finding $\lambda$ given that one has seen $n$ meteors.
This is the posterior probability described earlier 
which may be estimated using the theorem by Bayes:
\begin{equation}
P(\lambda|n) = \frac{ P(n|\lambda) \, P(\lambda) } {P(n)},
\end{equation}
where $P(n|\lambda)$ is the generative model,
i.e. the known probability to see $n$ meteors for a
given true rate $\lambda$. 
We may assume that meteors appear in a random way
following a Poissonian law 
for independent events:
\begin{equation}
P(n | \lambda) = \frac{\lambda^n e^{-\lambda}}{n!}.
\end{equation}
The function $P(n)$ serves to normalize the distribution to unity, while
$P(\lambda)$ expresses what we know about how
probable each of the possible true rates $\lambda$ are. This function is
called the prior and, ideally, should be a probability distribution
on its own. The challenge is to decide what we can 
assume about $P(\lambda)$ beforehand?

\section{Choice of the prior}
The criteria for choosing a suitable prior $P(\lambda)$ is a subject of 
debate in the statistical community. 
On one hand, one may decide to construct a prior based on previous evidence, 
for example the meteor activity in the past decade.
This is called an \emph{informative prior}.
On the other hand, one may prefer a prior which contains
only vague or general information and is not biased towards past observations.
This is called an \emph{objective prior}.

Whether or not it is appropriate to include previous observations 
in the computation of meteor rates is a philosophical question.
However, given the intrinsically variable nature 
of meteor showers, we suggest that an objective prior
is the only practical approach. 
We discuss the choice of such prior in what follows.

\subsection{Uniform prior}
A natural choice is a uniform prior $P(\lambda)$ = constant, i.e. all values of $\lambda$ are 
assumed to be equally likely. A uniform prior leads to a distribution which 
cannot be normalized, and is called an improper prior.
Any limit on ZHR, and be it very high, makes the distribution normalizable
though, that is, asymptotically a uniform prior is not a problem. 

Let us see what the uniform prior implies for the inferred rate.
The expectation value of the true meteor rate is defined as the 
sum of every possible value of $\lambda$ times its posterior probability. (In this paper,
we also use the terms ``average'' or ``mean'' as synonyms for expectation value.) 
For a continuous quantity such as $\lambda$, the expectation value is a normalized integral 
over all $\lambda$ which, for a uniform prior, reads:
\begin{eqnarray}
\E(\lambda)  &=& \int_0^\infty \! \lambda \, P(\lambda | n) \, \mathrm{d}\lambda \nonumber \\
&=& \int_0^\infty \! \lambda \, \frac{ \lambda^n }{ n! } \, e^{-\lambda} \, \mathrm{d}\lambda \nonumber \\
&=& (n+1) \int_0^\infty \! \frac{ \lambda^{n+1} }{ (n+1)! } \, e^{-\lambda} \, \mathrm{d}\lambda \nonumber \\
&=& n+1.
\label{integration}
\end{eqnarray}
Similarly, the error estimate for $\E(\lambda)$ follows from the mathematical definition of the variance,
often referred to as $\sigma^2$:
\begin{eqnarray}
\sigma^2(\lambda)  &=& \E[ (\lambda - \E(\lambda))^2 ] \nonumber \\
&=& \int_0^\infty \! P(\lambda | n) \, (\lambda - (n+1))^2  \, \mathrm{d}\lambda \nonumber \\
&=& \int\limits_0^\infty\frac{(n+1)(n+2)\lambda^{n+2}e^{-\lambda}}{(n+2)!} -\nonumber\\
&&         - \frac{2(n+1)^2\lambda^{n+1}e^{-\lambda}}{(n+1)!} +\nonumber\\
&&         + \frac{(n+1)^2\lambda^{n}e^{-\lambda}}{n!}d\lambda \nonumber\\
&=& (n+1)(n+2) - (n+1)^2\nonumber\\
&=& n+1.
\end{eqnarray}

The expectation value of the true activity rate is thus given by $(n+1)$ 
with spread of the distribution of $\sigma=\sqrt{n+1}$. These quantities 
are then simply divided by the weighted time $T$ to obtain the ZHR. 
This explains the formula given in the IMO Handbook.

\subsection{Exponential prior}
A uniform prior may not be ideal if we want to express the fact
that very large rates are very unlikely. Almost all rates ever
obtained are below say 1000~meteors per hour. A suitable
prior will decrease with rate, and it is mathematically convenient
to use power functions like
\begin{equation}
P(\lambda)=1/\lambda^{1-\alpha},
\label{prior}
\end{equation}
where $0\leq \alpha\leq 1$. Such a power-law has the advantage that
we can use $\Gamma$-functions for the derivation of $E(\lambda)$. 
The resulting expectation value now involves the prior and is
\begin{equation}
E(\lambda) = \frac{\int\limits_0^\infty \lambda \, P(\lambda)P(n|\lambda) d\lambda}
                  {\int\limits_0^\infty P(\lambda)P(n|\lambda)d\lambda},
\end{equation}
where the lower integral is to normalize the posterior distribution. 
Now, let's make use of the $\Gamma$-functions for which 
\begin{equation}
  n! = n\Gamma(n) = \Gamma(n+1) = \int\limits_0^\infty \lambda^n e^{-\lambda}d\lambda
\end{equation}
holds. We do not need to know how $\Gamma$ is computed, we only need
its properties to derive the expectation value:
\begin{eqnarray}
E(\lambda)&=&\frac{\int\limits_0^\infty\frac{1}{\lambda^{1-\alpha}}\frac{\lambda^n\,\lambda}{\Gamma(n+1)}e^{-\lambda}d\lambda}
                  {\int\limits_0^\infty\frac{1}{\lambda^{1-\alpha}}\frac{\lambda^n}{\Gamma(n+1)}e^{-\lambda}d\lambda}\nonumber\\
          &=&\frac{\frac{\Gamma(n+1+\alpha)}{\Gamma(n+1)}}
                  {\frac{\Gamma(n+\alpha)}{\Gamma(n+1)}}\nonumber\\
          &=&\frac{(n+\alpha)\Gamma(n+\alpha)}{\Gamma(n+\alpha)}\nonumber\\
          &=&n+\alpha.
\end{eqnarray}
Similarly, the variance is:
\begin{eqnarray}
\sigma^2(\lambda)  &=& E\left[\Bigl(\lambda-E(\lambda)\Bigr)^2\right]=\nonumber\\
&=&\frac{\int\limits_0^\infty P(\lambda)P(n|\lambda)\Bigl(\lambda-(n+\alpha)\Bigr)^2 d\lambda}
       {\int\limits_0^\infty P(\lambda)P(n|\lambda)d\lambda}\nonumber\\
&=&\frac{(n+1+\alpha)(n+\alpha)\Gamma(n+\alpha)}
        {\Gamma(n+\alpha)}-\nonumber\\
&&\phantom{=}-\frac{2(n+\alpha)^2\Gamma(n+\alpha)+(n+\alpha)^2\Gamma(n+\alpha)}
        {\Gamma(n+\alpha)}\nonumber\\
&=&(n+1+\alpha)(n+\alpha) - 2(n+\alpha)^2 + (n+\alpha)^2\nonumber\\
&=&n+\alpha,
\end{eqnarray}

The expectation value of the true activity rate is thus given by $(n+\alpha)$ 
with a spread $\sigma=\sqrt{n+\alpha}$, for a given prior $1/\lambda^{1-\alpha}$.
Again, these priors are normalizable for $0\leq\alpha\leq1$
by enforcing a limitation of the valid range.

One question remains: which is the most appropriate value to adopt for $\alpha$?
On one hand, the uniform prior ($\alpha=1$) yields the expectation value $n+1$, which
is likely to overestimate the meteor rate due to assumption that any arbitrarily large rate
is equally likely as the zero rate. On the other hand,
the prior $1/\lambda$ ($\alpha=0$) yields a ``traditional'' extrapolation $E(\lambda) = n$,
which is likely to underestimate the rate as explained previously.
It appears appropriate to adopt a prior somewhere inbetween $0 < \alpha < 1$.

\subsection{Jeffreys prior; $\alpha$ = 0.5}
The problem of choosing a suitable prior for a Poissonian process
exists in other fields (e.g. radioactive decay counts, 
neutrino detections).
An axiomatic solution has previously
been proposed by Harold Jeffreys.
He required that a prior should be ``invariant under reparameterization'',
i.e. a prior should not depend on the variable investigated (Jeffreys 1946, 1961;
Kass \& Wasserman 1996). One could, for example, be 
interested in the rate $\lambda$ as well as in the mean time between 
events $\mu = 1/\lambda$. The priors for both expectation values
should be compatible. For general relations between different 
quantities, the requirement of compatibility is written mathematically 
as
\begin{equation}
  P(\lambda) d\lambda = P(\mu) d\mu.
\end{equation}
For a Poissonian distribution, such a general compatibility is achieved 
for a prior $\alpha=0.5$. In this case, the estimate is not specific to either computing 
the rate or the mean time lapse or something else. 

Because there are no further statistical principles to decide which
prior is to be preferred, we suggest the use of Jeffreys prior
\begin{equation}
P(\lambda)=1/\lambda^{0.5},
\end{equation}
for future estimates of the rate. The expectation value for the meteor rate
is then
\begin{equation}
E(\lambda) = (n+0.5) \pm \sqrt{n+0.5}.
\end{equation}
An illustration of the posterior distribution $P(\lambda|n)$ under the assumption
of Jeffreys' prior is shown in Figure~1 for different values of $n$.

Note that in most cases, when $n$ is large ($n \gg 0.5$), the differences between
the various priors are negligable.

\begin{figure*}[t]
\centering
\includegraphics[width=0.6\textwidth]{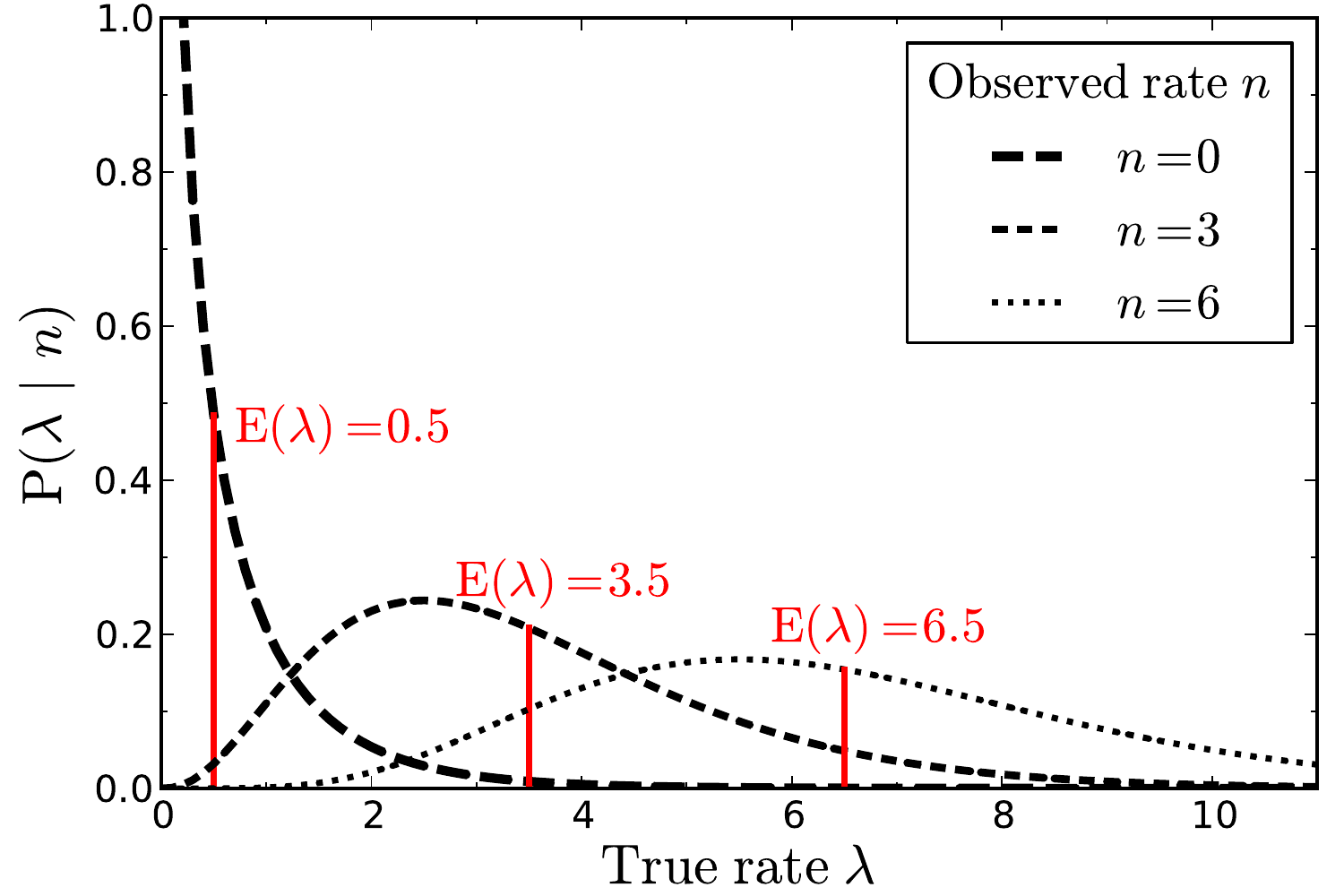}
\caption{Posterior distribution $P(\lambda|n)$ of the true meteor rate
under Jeffreys' prior ($\alpha=0.5$), plotted for different observed rates ($n=0,3,6$).
Vertical lines indicate the position of the expectation values $\E(\lambda)=n+0.5$.}
\end{figure*}

\section{Confidence intervals}
The standard deviation $\sigma = \sqrt{n + 0.5}$ characterizes the spread in the posterior distribution around the expectation value.
In the case of a Gaussian distribution, the standard deviation corresponds to a confidence interval
(i.e. the 68\%-confidence interval of a Gaussian distribution is located between $-1\sigma$ and $+1\sigma$ from the mean).

However, the posterior $P(\lambda|n)$ does not follow a Gaussian shape and is more similar to a Poissonian
distribution (though a Gaussian shape is approached for large $n$).
The posterior is asymmetric with a tail towards high meteor rates,
which makes it somewhat misleading to characterize the uncertainty with a single number.
A better way to characterize the uncertainty is to compute the (asymmetric) error margins 
of the 68\%-confidence interval. This may be done as follows.

Given the posterior distribution
\begin{equation}
  P(\lambda|n) = \frac{\lambda^{n-1+\alpha} e^{-\lambda}}{\Gamma(n+\alpha)}
\end{equation}
The corresponding cumulative distribution function is
\begin{eqnarray}
P'(\lambda\leq \lambda_{\rm marg} | n ) &=& 
\int_0^{\lambda_{\rm marg}} \frac{ t^{n-1+\alpha} e^{-t} } { \Gamma(n+\alpha) } dt\nonumber\\
&=& 1 - \frac{ \Gamma(\alpha + n, \lambda_{\rm marg}) } { \Gamma(\alpha + n) },
\end{eqnarray}
with the incomplete $\Gamma$ function which for integers $n>0$ can be computed as
\begin{equation}
\Gamma(n,\lambda_{\rm marg}) = (n-1)! \, e^{-\lambda_{\rm marg}} \sum_{k=0}^{n-1} \frac{\lambda_{\rm marg}^k}{k!}
\end{equation}
By integrating the cumulative distribution function numerically for different probabilities
($P' = 2.5$\%, $16$\%, $84$\%, and $97.5$\%), we obtain useful quantiles which correspond to
the central 68\%- and 95\%-confidence intervals.

These quantiles are shown in Table~1, given as a multiplier of the true rate.
For example, after computing the ZHR, one may look up the corresponding relative margins for a given $n_{\rm tot}$
in Table~1 and obtain the 68\%-interval by computing
${\rm ZHR}\cdot\delta_{\rm 68, low}$ (negative margin) and ${\rm ZHR}\cdot\delta_{\rm 68, high}$ (positive margin).

\begin{table}
\caption{Margins of the 68\%- and 95\%-confidence intervals,
given as multipliers to the ZHR. 
After computing ${\rm ZHR}=(n_{\rm tot}+0.5)/T$, the
$\pm$-values can be obtained by computing ${\rm ZHR}\cdot\delta_{\rm 68, low}$ (negative margin)
and ${\rm ZHR}\cdot\delta_{\rm 68, high}$ (positive margin) for the appropriate $n_{\rm tot}$.
\label{errors}}
\begin{tabular}{rcccc}
\hline
$n_{\rm tot}$& $\delta_{\rm 95, low}$&$\delta_{\rm 68, low}$ &   $\delta_{\rm 68, high}$&$\delta_{\rm 95, high}$\\
\hline
  0 &         -1.00 &         {\bf -0.96} &         {\bf +0.99} &         +4.02\\
  1 &         -0.93 &         {\bf -0.72} &         {\bf +0.73} &         +2.12\\
  2 &         -0.83 &         {\bf -0.59} &         {\bf +0.59} &         +1.57\\
  3 &         -0.76 &         \multicolumn{2}{c}{  {\bf $\pm$0.51}} &         +1.29\\
  4 &         -0.70 &         \multicolumn{2}{c}{  {\bf $\pm$0.45}} &         +1.11\\
  5 &         -0.65 &         \multicolumn{2}{c}{  {\bf $\pm$0.41}} &         +0.99\\
  6 &         -0.61 &         \multicolumn{2}{c}{  {\bf $\pm$0.38}} &         +0.90\\
  7 &         -0.58 &         \multicolumn{2}{c}{  {\bf $\pm$0.36}} &         +0.83\\
  8 &         -0.56 &         \multicolumn{2}{c}{  {\bf $\pm$0.34}} &         +0.78\\
  9 &         -0.53 &         \multicolumn{2}{c}{  {\bf $\pm$0.32}} &         +0.73\\
 10 &         -0.51 &         \multicolumn{2}{c}{  {\bf $\pm$0.30}} &         +0.69\\
 11 &         -0.49 &         \multicolumn{2}{c}{  {\bf $\pm$0.29}} &         +0.66\\
 12 &         -0.48 &         \multicolumn{2}{c}{  {\bf $\pm$0.28}} &         +0.63\\
 13 &         -0.46 &         \multicolumn{2}{c}{  {\bf $\pm$0.27}} &         +0.60\\
 14 &         -0.45 &         \multicolumn{2}{c}{  {\bf $\pm$0.26}} &         +0.58\\
 15 &         -0.43 &         \multicolumn{2}{c}{  {\bf $\pm$0.25}} &         +0.56\\
 16 &         -0.42 &         \multicolumn{2}{c}{  {\bf $\pm$0.24}} &         +0.54\\
 17 &         -0.41 &         \multicolumn{2}{c}{  {\bf $\pm$0.24}} &         +0.52\\
 18 &         -0.40 &         \multicolumn{2}{c}{  {\bf $\pm$0.23}} &         +0.50\\
 19 &         -0.39 &         \multicolumn{2}{c}{  {\bf $\pm$0.22}} &         +0.49\\
 20 &         -0.39 &         \multicolumn{2}{c}{  {\bf $\pm$0.22}} &         +0.48\\
 22 &         -0.37 &         \multicolumn{2}{c}{  {\bf $\pm$0.21}} &         +0.45\\
 24 &         -0.36 &         \multicolumn{2}{c}{  {\bf $\pm$0.20}} &         +0.43\\
 26 &         -0.34 &         \multicolumn{2}{c}{  {\bf $\pm$0.19}} &         +0.42\\
 28 &         -0.33 &         \multicolumn{2}{c}{  {\bf $\pm$0.19}} &         +0.40\\
 30 &         -0.32 &         \multicolumn{2}{c}{  {\bf $\pm$0.18}} &         +0.38\\
\hline
\end{tabular}
\end{table}

It is interesting to note that the 68\% interval approaches symmetry from $n_{\rm tot} \gtrsim 3$,
while the 95\% interval is more sensitive to the wings 
and remains asymmetric even beyond $n_{\rm tot} \gg 1000$.

For $n_{\rm tot}$ larger than about 30, the 68\%-interval approaches a Gaussian shape
and the margins can be approximated using $\sigma = \sqrt{n_{\rm tot}+0.5} / T$ 
($= \rm ZHR / \sqrt{n_{\rm tot}+0.5}$).

Finally, we remind the reader that the margins in Table~1 only represent the uncertainty which is due to Poissonian statistics. The true uncertainty of a ZHR estimate is likely to be somewhat larger due to observing errors (e.g., uncertainty in the limiting magnitude determination). Fortunately, the impact of such observing errors is likely to be small when data from a sufficient number of independent observers is averaged.

\section{Examples}
Now let us consider the formula using a few examples. 
If one meteor was seen in five minutes, 
the rate equals $E({\rm ZHR}) = \frac{n+0.5}{T} = 18.0_{-13.0}^{+13.1}$.
When zero meteors are seen during five minutes, the rate equals ${\rm ZHR} = 6.0_{-5.8}^{+5.9}$.
The error margins are actually so close to symmetric that we can always give a
single value for the error bars, i.e.\ ${\rm ZHR} = 18\pm 13$ and ${\rm ZHR} = 6\pm6$
respectively. Although the rounded margins suggest so, the lower margin is not zero!

If zero meteors were seen in four hours, the rate equals to ${\rm ZHR} = 0.125_{-0.120}^{+0.124}$ 
Indeed, rates can only be constrained to values close to zero when no meteors are observed 
for a very long period. A more ``normal'' case for a minor shower would be say a total of 
12~meteors in a total of 4~hours, delivering ${\rm ZHR}=3.1\pm0.9$. A larger
number of meteors of say 34~meteors in 11~hours gives simply ${\rm ZHR}=3.1\pm0.5$
with the above error for large meteor numbers.

\section{The influence of correction factors}
In the previous sections we have ignored the specific correction factors which
are used to obtain a standardized ZHR (e.g. to account for limiting magnitude and radiant elevation).
Given two rates; one without correction, $\hat\lambda$ and one with correction, $\lambda$, 
we have assumed there is a factor $f$ which does {\em not\/} depend on the rate:
\begin{equation}
  f\,\lambda = \hat\lambda
\end{equation}
Indeed, for a set of $N$ observing periods being combined in one expectation
value for $\lambda$, all having different correction factors $f_i$, we obtain
\begin{eqnarray}
E(\lambda)&\!\!=\!\!&\frac{\int\limits_0^\infty\frac{1}{\lambda^{1-\alpha}}\lambda\frac{(f_1\lambda)^{n_1}(f_2\lambda)^{n_2}\cdots(f_N\lambda)^{n_N}}
             {n_1!n_2!\cdots n_N!} e^{-\sum f_i\lambda}d\lambda}
                  {\int\limits_0^\infty\frac{1}{\lambda^{1-\alpha}}\frac{(f_1\lambda)^{n_1}(f_2\lambda)^{n_2}\cdots(f_N\lambda)^{n_N}}
             {n_1!n_2!\cdots n_N!} e^{-\sum f_i\lambda}d\lambda}\nonumber\\
          &\!\!=\!\!&\frac{\Gamma(n_{\rm tot}+1+\alpha)}{\left(\sum f_i\right)^{n_{\rm tot}+1+\alpha}}
              \frac{\left(\sum f_i\right)^{n_{\rm tot}+\alpha}}{\Gamma(n_{\rm tot}+\alpha)}\nonumber\\
          &\!\!=\!\!&\frac{n_{\rm tot}+\alpha}{\sum f_i}.
\end{eqnarray}
In other words, our method may be applied regardless of the
values of the correction factors.

\section{Discussion and conclusion}
In conclusion, we recommend to compute ZHR values using the term $n+0.5$:
\begin{equation}
\E(\ZHR) = \frac{\left(n_{\rm tot}+0.5\right) \, r^{6.5-{\rm lm}}}{T_{\rm eff}\sin h_{\rm R}},
\end{equation}
with error margins:
\begin{equation}
\Delta{\rm ZHR} = \frac{\rm ZHR}{\sqrt{n_{\rm tot}+0.5}}.
\label{finalerror}
\end{equation}
Note that for $n_{\rm tot}\leq 30$, the error margins listed in Table~1 should be used instead of Eqn.~\ref{finalerror}
to obtain a 68\%-confidence interval.

This method of computing the ZHR adds a small bias taking 
into account the asymmetry of possible rates. In particular, it is essential 
to adopt the method whenever rates are based on less than $\sim10$ meteors. Such 
situations commonly occur when a major shower is analysed using very short 
(e.g.\ 1-minute) intervals. When computing a rate based on a number of observing
periods (indexed with $i$), {\em never ever\/} compute the rate from $n_i+0.5$ 
for individual periods and average them afterwards. A more accurate estimate is based on the sum of 
meteors from these observing periods, and hence on $n_{\rm tot}+0.5$. Finally, 
it is, of course, much better to have enough observations and large enough $n_{\rm tot}$ 
that the subtleties of choosing a good prior are no longer important, i.e.\ 
$n_{\rm tot} \gg 0.5$.

Comprehensive information about Bayesian inference in general and
the choice of priors can be found in e.g.\ Kass \&Wasserman (1996)
and Bolstad (2007; Chapter~10 for priors).

\end{WGNpaper}
\end{document}